\documentclass{article}

\usepackage{arxiv}

\usepackage[utf8]{inputenc} 
\usepackage[T1]{fontenc}    
\usepackage{hyperref}       
\usepackage{url}            
\usepackage{booktabs}       
\usepackage{amsfonts}       
\usepackage{nicefrac}       
\usepackage{microtype}      
\usepackage{lipsum}		
\usepackage{graphicx}
\usepackage{natbib}
\usepackage{doi}
\usepackage{amsmath}

\usepackage{fancyhdr}
\usepackage{textcomp}

\title{From Craft Practice to Aesthetic Cognition Transmission: Workflow Cognition Translation for AI-native Intangible Cultural Heritage Education}

\author{ 
	\href{https://orcid.org/0009-0004-1760-0149}{\includegraphics[scale=0.06]{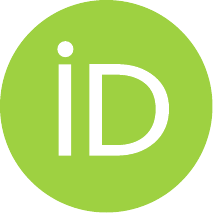}\hspace{1mm}Annie Yuan} \\
	School of Computer Science\\
	The University of Sydney\\
	NSW, 2006, Australia \\
	\texttt{annie.yuan@sydney.edu.au} \\
}

\renewcommand{\shorttitle}{From Craft to Aesthetic Cognition Transmission}

\hypersetup{
pdftitle={Tacit Signal Infrastructure: Towards AI Systems that Model Expert Sensing Over Time},
pdfsubject={Artificial Intelligence, Human--AI Interaction, Expert Cognition, Cognitive Infrastructure},
pdfauthor={Annie Yuan},
pdfkeywords={Tacit Signal Infrastructure, Expert Weak Signal Perception, Weak Cognitive Signals, Tacit Signal Computability, Longitudinal Cognitive Operations, Expert Cognition, Weak Signals}
}

\begin{document}
\maketitle
\thispagestyle{plain}

\begin{abstract}
Intangible Cultural Heritage (ICH) education has traditionally relied on apprenticeship, embodied participation, and long-term engagement with masters, materials, and cultural environments. While these modes of transmission remain essential, they are difficult to scale. Existing digital heritage initiatives have expanded documentation and access, but often preserve artefacts, procedures, and representations of practice rather than the aesthetic and cognitive structures through which expertise operates.
This paper argues that the future challenge of ICH education is not only the transmission of craft techniques, but the scalable transmission of aesthetic cognition: the perception, judgement, interpretation, and culturally situated meaning-making through which aesthetic expertise develops. Drawing on aesthetic education, tacit knowledge, cognitive apprenticeship, and expert cognition, we propose a shift from craft transmission to \textit{Aesthetic Cognition Transmission}.
To support this shift, we introduce \textit{Workflow Cognition} as a model of how experts coordinate perception, judgement, decision-making, and action within evolving workflows. We then propose \textit{Workflow Cognition Translation} as a methodological framework for transforming expert workflow cognition into computable educational representations for AI-native learning systems.
The paper makes three contributions: it reframes ICH education around aesthetic cognition transmission; introduces Workflow Cognition Translation as a method for representing expert aesthetic cognition; and outlines an AI-native cognitive apprenticeship infrastructure involving AI Expert Twins, workflow-based tutoring, and progressive learner participation. Rather than replacing masters, workshops, or embodied practice, the framework positions AI as a cognition mediation infrastructure for expanding access to heritage expertise.
\end{abstract}

\keywords{
Intangible Cultural Heritage \and
Aesthetic Cognition \and
Aesthetic Cognition Transmission \and
Workflow Cognition \and
Workflow Cognition Translation \and
AI-native Education \and
Cognitive Apprenticeship \and
Digital Heritage
}

\section{Introduction}
Intangible Cultural Heritage (ICH) encompasses a wide range of cultural practices, skills, performances, and traditions that are transmitted across generations through participation in communities of practice. Historically, the development of expertise within many heritage domains has depended upon apprenticeship, observation, embodied participation, and prolonged engagement with materials, tools, and cultural environments. Whilst these approaches remain central to heritage preservation, they also create challenges for accessibility and scalability. Many forms of heritage expertise require direct access to masters, specialised workshop environments, and years of sustained practice before meaningful participation becomes possible.

Digital technologies have substantially improved the documentation and preservation of heritage knowledge through archives, digital collections, multimedia records, virtual exhibitions, and interactive cultural experiences. However, much of this work has focused on preserving artefacts, procedures, and representations of practice rather than the cognitive structures through which expertise is enacted. As a result, access to information about heritage practices has expanded considerably, whilst access to the underlying mechanisms of expert perception, judgement, interpretation, and decision-making remains comparatively limited.

This limitation reflects a broader challenge in both heritage education and expertise research. Educational traditions associated with aesthetic education have long argued that expertise involves more than the acquisition of procedural knowledge or technical skill. Expert practitioners develop specialised ways of perceiving, evaluating, interpreting, and responding to situations that emerge through extended participation within a domain. Similarly, research on tacit knowledge, situated learning, cognitive apprenticeship, and expert performance has emphasised that much of what distinguishes experts from novices cannot be fully captured through explicit instruction alone. Expertise often resides in perceptual sensitivities, judgement processes, contextual reasoning, and adaptive responses that develop gradually through practice.

These observations raise an important question for the future of AI-mediated heritage learning:

\begin{quote}
How can the aesthetic and cognitive structures of expert heritage practice be represented and transmitted in ways that support scalable participation, complementing apprenticeship, embodied practice, and cultural context?
\end{quote}

This paper argues that answering this question requires a shift in educational perspective. Rather than viewing heritage education primarily as the transmission of craft techniques, we propose understanding it as the transmission of \textit{aesthetic cognition}. We define aesthetic cognition as the integrated system of perception, interpretation, judgement, evaluation, and culturally situated meaning-making through which aesthetic expertise operates. From this perspective, the primary educational challenge is not simply enabling learners to reproduce artefacts, but enabling participation in expert ways of seeing, judging, and acting.

To support this shift, we introduce the concept of \textit{Workflow Cognition}. Workflow Cognition conceptualises expertise as a dynamic cognitive system through which experts coordinate perception, interpretation, judgement, decision-making, and action within evolving workflows. Building upon this foundation, we propose \textit{Workflow Cognition Translation}, a methodological framework for transforming expert workflow cognition into computable educational representations that can support AI-native learning systems. Rather than digitising instructional content alone, Workflow Cognition Translation seeks to represent the cognitive mechanisms through which expertise is enacted, adapted, and refined in practice.

The paper makes three contributions. First, it proposes \textit{Aesthetic Cognition Transmission} as a new educational paradigm for ICH learning, positioning aesthetic expertise rather than craft reproduction as the central object of transmission. Second, it introduces Workflow Cognition and Workflow Cognition Translation as conceptual and methodological foundations for representing expert cognition within AI-native educational environments. Third, it outlines an AI-native cognitive apprenticeship infrastructure composed of AI Expert Twins, workflow-based tutoring systems, and progressive learner participation pathways designed to expand access to heritage expertise whilst preserving the importance of masters, workshops, and embodied practice.

The remainder of the paper develops this argument in five stages. We first situate aesthetic cognition within traditions of aesthetic education and expertise development. We then introduce Workflow Cognition as an ontology of expert practice and Workflow Cognition Translation as a computational methodology for representing expertise. Building upon these foundations, we describe how translated workflow cognition can support AI-native cognitive apprenticeship systems and scalable participation in heritage learning. Finally, we discuss broader implications for AI-native education, cultural preservation, and the future representation of human expertise.

\section{Aesthetic Cognition and Heritage Education}

This section situates the paper within three relevant bodies of work: aesthetic education, tacit expertise, and cognitive apprenticeship. These traditions provide the theoretical basis for treating heritage education not merely as the reproduction of craft techniques, but as the development of expert ways of perceiving, judging, interpreting, and acting. Building on this foundation, the section introduces \textit{aesthetic cognition} as the central educational object of AI-native ICH education.

\subsection{Aesthetic Education Beyond Skill Acquisition}

Educational theories have long argued that learning involves more than the acquisition of knowledge and technical skill. Within traditions of aesthetic education, scholars have emphasised perception, judgement, interpretation, imagination, embodied experience, and meaning-making as central dimensions of learning. Dewey's account of art as experience positions aesthetic understanding as an active process that emerges through engagement with materials, environments, and lived experience rather than through passive reception of information \cite{dewey2024art}. His broader theory of experience and education further suggests that learning should be understood as an active, situated, and developmental process rather than the mere accumulation of instructional content \cite{dewey1986experience}.

Eisner similarly argues that the arts cultivate forms of perception and judgement that are central to human understanding. In his account, education should develop learners' capacities to notice, discriminate, and make qualitative judgements about features of experience that may be invisible to novices \cite{eisner2003arts}. His work on educational connoisseurship and criticism further emphasises that expert understanding often involves refined perception and interpretive judgement rather than rule-based evaluation alone \cite{eisner2017enlightened}. Greene also situates aesthetic education in relation to imagination, meaning-making, and enlarged forms of human awareness, suggesting that aesthetic learning can open new ways of perceiving and engaging with the world \cite{greene2000releasing}. Shusterman's pragmatist aesthetics and somaesthetics further extend this view by emphasising the embodied dimensions of aesthetic experience and the role of bodily awareness in perception and judgement \cite{shusterman2000pragmatist, shusterman2008body}.

From this perspective, expertise cannot be reduced to procedural competence alone. Experts develop specialised ways of seeing, interpreting, and evaluating situations that guide their actions and decisions. Learning therefore involves not only knowing how to perform a task, but also developing the perceptual, interpretive, and evaluative capacities that allow meaningful participation within a practice.

These ideas are particularly relevant to Intangible Cultural Heritage (ICH). While heritage education is often associated with the transmission of craft techniques, many heritage traditions embody forms of aesthetic understanding that extend beyond technical execution. Practices such as jade carving, ceramics, weaving, calligraphy, ritual performance, and martial arts require practitioners to develop aesthetic judgement, cultural interpretation, material sensitivity, embodied awareness, and contextual understanding. The educational significance of these practices lies not only in what learners produce, but also in how they learn to perceive, interpret, and engage with cultural forms.

\subsection{Tacit Knowledge, Expertise, and Cognitive Apprenticeship}

The importance of perception and judgement in expertise has been recognised across several research traditions. Polanyi's concept of tacit knowledge highlights that experts often possess forms of understanding that cannot be fully articulated through explicit rules or verbal descriptions \cite{polanyi2009tacit}. Expert performance frequently depends on subtle forms of perception, recognition, anticipation, and judgement that emerge through prolonged experience rather than formal instruction.

Situated learning theory similarly suggests that learning is not simply the internal acquisition of abstract knowledge, but participation in socially and culturally organised practices. Lave and Wenger's account of legitimate peripheral participation explains how learners gradually enter communities of practice through observation, participation, and increasing responsibility \cite{lave1991situated}. Wenger's later work further emphasises that learning, meaning, and identity are formed through participation in communities of practice rather than through decontextualised instruction alone \cite{wenger1999communities}. These perspectives are highly relevant to ICH education, where expertise is often transmitted through sustained participation in workshops, communities, and culturally embedded environments.

Research on expertise also shows that experts differ from novices not simply because they possess more information, but because they perceive situations differently, recognise meaningful patterns, and organise action around domain-specific structures \cite{chi2014nature, ericsson2018cambridge}. Ericsson's work on deliberate practice emphasises that expertise develops through sustained, structured, and feedback-rich engagement over time \cite{ericsson1993role}. Simon's work on artificial and designed systems provides a broader foundation for understanding expertise as structured problem-solving within complex environments \cite{simon2019sciences}. Klein's naturalistic account of expert decision-making further shows that experts often rely on situation assessment, pattern recognition, and experience-based judgement under real-world constraints rather than explicit rule application alone \cite{klein2017sources}.

Within educational theory, cognitive apprenticeship was proposed as a response to the challenge of making expert thinking visible to learners. Collins, Brown, and Newman argue that traditional apprenticeship allows learners to observe expert practice, but that many aspects of expert reasoning remain hidden \cite{collins2018cognitive}. Cognitive apprenticeship therefore seeks to externalise expert thinking through modelling, coaching, scaffolding, articulation, reflection, and exploration. The related formulation of cognitive apprenticeship as ``making thinking visible'' is especially relevant to heritage education, where the most important dimensions of expertise may be embedded in tacit perception, judgement, and contextual reasoning \cite{collins1991cognitive}. Brown, Collins, and Duguid similarly argue that knowledge is situated within the activities, cultures, and contexts in which it is used \cite{brown1989situated}. Rogoff's account of apprenticeship in thinking further supports the view that cognitive development occurs through guided participation in socially organised activity \cite{rogoff1990apprenticeship}.

These perspectives suggest that expertise should be understood not merely as knowledge possession, but as participation in systems of perception, judgement, interpretation, and action. This observation is particularly significant for heritage education, where many of the most valuable aspects of expertise are embedded within aesthetic judgement, cultural understanding, embodied awareness, and material sensitivity rather than explicit procedural knowledge.

\subsection{From Craft Transmission to Aesthetic Cognition Transmission}

Traditional approaches to heritage education have largely focused on craft transmission. Within such models, educational success is often measured by the learner's ability to reproduce techniques, procedures, and culturally recognised artefacts. While this remains important for safeguarding endangered traditions, it captures only part of what is transmitted through heritage practice.

What distinguishes experts is often not the execution of a technique itself, but the cognitive processes through which techniques are selected, adapted, interpreted, and evaluated. Two practitioners may possess similar technical skills yet arrive at different outcomes because they perceive different possibilities, make different aesthetic judgements, or interpret cultural meanings differently. The educational challenge is therefore not solely the transmission of craft procedures, but the transmission of the underlying cognitive structures that support expert practice.

This paper refers to these structures as \textit{aesthetic cognition}. We define aesthetic cognition as the integrated system of perception, interpretation, judgement, evaluation, and culturally situated meaning-making through which aesthetic expertise operates. Aesthetic cognition enables practitioners to recognise patterns, evaluate alternatives, interpret cultural significance, respond to contextual constraints, and make decisions within a domain of practice.

From this perspective, the primary object of transmission in heritage education is not the artefact itself, nor the procedural steps used to create it. Rather, it is participation in the aesthetic cognition through which heritage expertise is enacted and developed. We therefore propose \textit{Aesthetic Cognition Transmission} as an educational paradigm that shifts the focus of heritage learning from craft reproduction towards participation in expert ways of perceiving, judging, and acting.

\begin{figure}[htbp]
    \centering
    \includegraphics[width=0.95\linewidth]{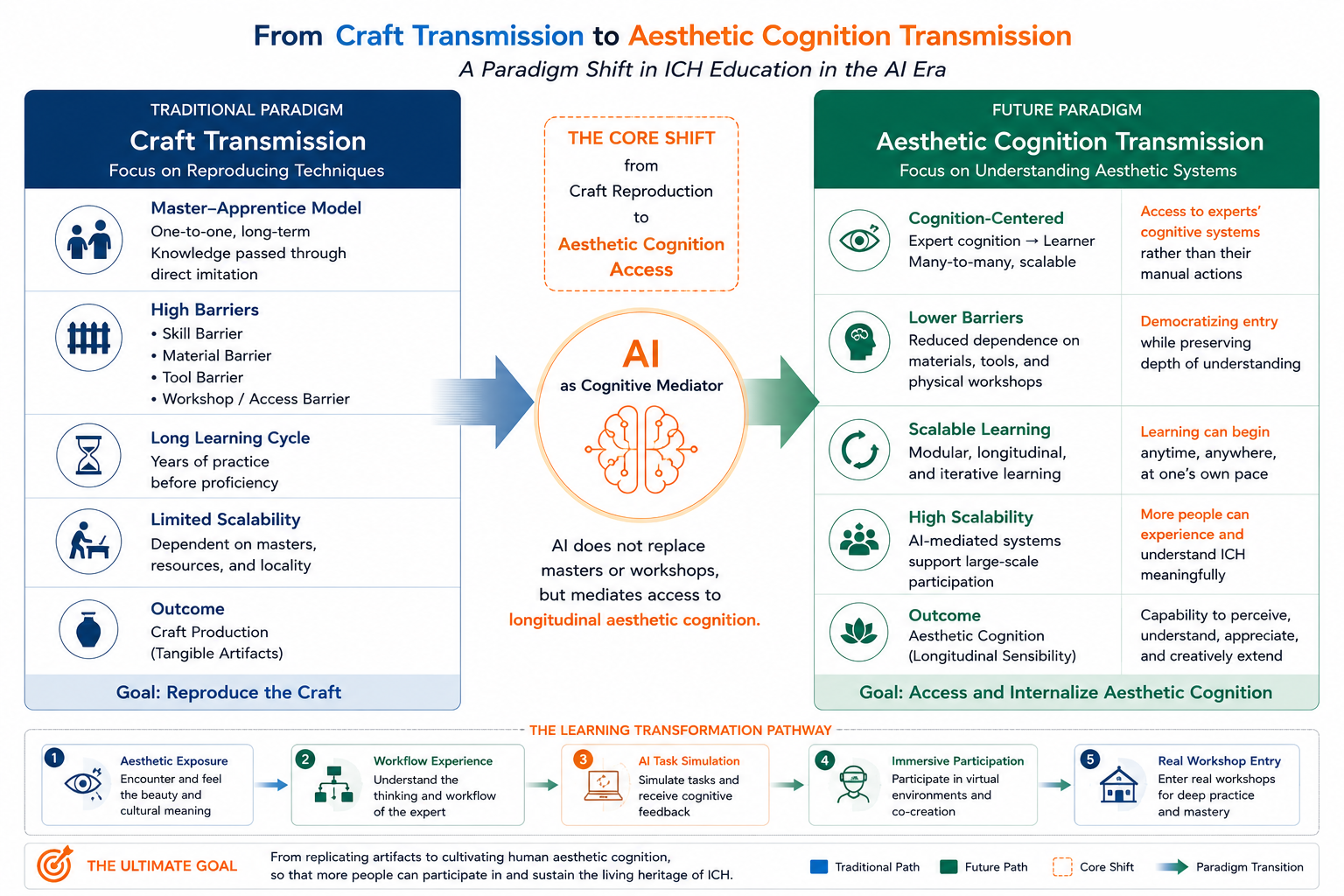}
    \caption{From Craft Transmission to Aesthetic Cognition Transmission: A paradigm shift in ICH education in the AI era.}
    \label{fig:craft-transmission}
\end{figure}

Figure~\ref{fig:craft-transmission} illustrates this paradigm shift. Rather than treating craft reproduction as the central educational objective, the framework positions aesthetic cognition as the primary object of transmission and AI-mediated learning as a mechanism for expanding participation in heritage expertise.

\subsection{The Structure of Aesthetic Cognition}

Having established aesthetic cognition as the educational object of transmission, a further question emerges: what constitutes aesthetic cognition itself? While aesthetic cognition manifests differently across heritage domains, three recurring dimensions can be identified across a wide range of practices: image, behaviour, and language.

\textit{Image} refers to visual forms, symbolic structures, compositional patterns, material aesthetics, and culturally meaningful representations embedded within heritage artefacts and practices. Through image systems, practitioners learn to recognise aesthetic relationships, visual hierarchies, stylistic conventions, and symbolic meanings.

\textit{Behaviour} refers to embodied participation in workflow. Heritage expertise is not acquired solely through observation but through engagement with procedural rhythms, material interactions, operational sequences, and behavioural coordination. Behaviour therefore represents the embodied dimension of aesthetic cognition.

\textit{Language} refers to the semantic systems through which expertise is communicated, interpreted, and transmitted. Heritage communities often employ specialised vocabularies, metaphors, narratives, evaluative expressions, and culturally situated concepts that encode forms of understanding accumulated through generations of practice.

Together, image, behaviour, and language form an integrated system through which aesthetic understanding is generated, communicated, and refined. Rather than functioning independently, these dimensions continuously interact to shape perception, judgement, action, and cultural participation.

\begin{figure}[htbp]
    \centering
    \includegraphics[width=0.95\linewidth]{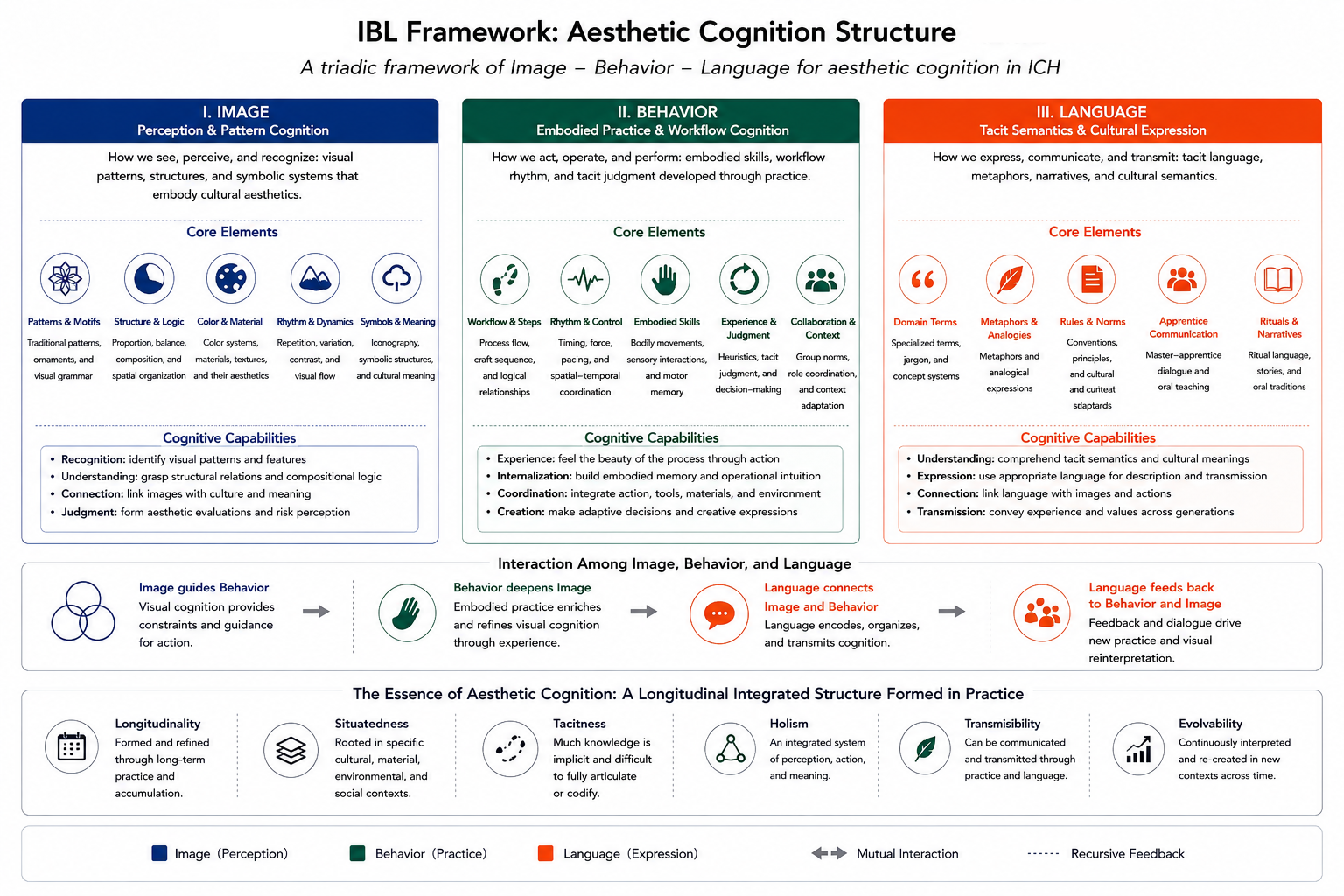}
    \caption{IBL Framework: A triadic model of Image, Behaviour, and Language in aesthetic cognition.}
    \label{fig:ibl-framework}
\end{figure}

To conceptualise this relationship, we introduce the \textit{IBL Framework} (Image--Behaviour--Language). As shown in Figure~\ref{fig:ibl-framework}, the framework provides an initial representational model for understanding the internal structure of aesthetic cognition within heritage practice.

This reframing does not diminish the importance of apprenticeship, workshops, or embodied practice. Instead, it clarifies what these educational forms transmit and provides a foundation for exploring how aspects of aesthetic expertise may become accessible within AI-mediated learning environments. If aesthetic cognition is the educational object, a subsequent challenge emerges: how can such cognition be represented, analysed, and translated into forms capable of supporting scalable participation? The following section addresses this question through the concept of Workflow Cognition.

\section{Workflow Cognition as an Ontology of Expertise}

Section 2 argued that the central educational object of AI-native ICH education is not craft reproduction alone, but aesthetic cognition: the expert capacity to perceive, judge, interpret, and act within culturally situated practice. This section develops the next step in the framework by introducing \textit{Workflow Cognition}. While aesthetic cognition defines what heritage education seeks to transmit, Workflow Cognition explains how such cognition operates within expert practice.

\subsection{From Tacit Knowledge to Workflow Cognition}

The concept of tacit knowledge is useful for explaining why expert practice cannot be fully captured through explicit instruction. However, tacit knowledge alone does not explain how expertise unfolds during action. Experts do not simply possess hidden knowledge; they continuously coordinate perception, judgement, decision-making, and action while responding to changing materials, constraints, and opportunities.

This distinction is important for ICH education. A master craftsperson does not merely follow a stored procedure. During practice, the expert attends to subtle cues, interprets material conditions, evaluates emerging forms, anticipates possible outcomes, and adapts the workflow accordingly. Expertise therefore appears not as a static body of knowledge, but as an active and evolving cognitive process embedded within practice.

We use the term \textit{Workflow Cognition} to describe this process. It provides an ontology of expertise that represents not only what experts know, but how expertise operates, adapts, and develops within real-world workflows.

\begin{quote}
\textbf{Ontology Definition.} \textit{Workflow Cognition} is the underlying cognitive architecture through which experts organise perception, contextual interpretation, judgement, decision-making, embodied action, and adaptive reasoning into coherent workflow evolution across time.
\end{quote}

This definition positions expertise as a process-oriented structure rather than a static repository of knowledge. From this perspective, expertise is not defined primarily by the quantity of information an individual possesses, but by the cognitive structures that enable perception, judgement, and action to operate coherently under changing conditions.

\begin{quote}
\textbf{Mechanism Definition.} \textit{Workflow Cognition} operates through dynamic cognitive structures that continuously transform perception into judgement, judgement into action, and action into adaptive workflow evolution under real-world constraints.
\end{quote}

The mechanism definition emphasises that expertise is not a fixed state. It is an active process of cognitive coordination in which experts interpret information, evaluate possibilities, anticipate consequences, modify actions, and adjust workflows in response to environmental and material feedback. Workflow Cognition therefore represents not simply what experts know, but how expertise continuously operates, adapts, and evolves within real-world practice.

\subsection{Thinking Flow, Workflow, and Artefact}

Workflow Cognition can be understood as a dynamic coupling system across three layers: the cognitive layer, the behavioural layer, and the material layer. The first layer is \textit{thinking flow}, which refers to the ongoing cognitive processes through which experts perceive situations, interpret context, evaluate alternatives, anticipate consequences, and generate decisions. These processes are often only partially verbalised and may remain invisible to external observers.

The second layer is \textit{workflow}, which refers to the observable sequence of actions, operations, gestures, procedures, and material interactions through which practice is enacted. In heritage education, workflow is usually easier to document because it can be demonstrated, filmed, described, or imitated. However, workflow alone does not explain why experts make particular decisions, recognise specific opportunities, or adapt when unexpected conditions emerge.

The third layer is the \textit{artefact layer}, or material layer, which refers to the evolving material object, visual form, surface qualities, structural constraints, and cultural meanings that emerge through practice. The artefact is not merely the final output of expert action. It is also a source of feedback that continuously reshapes perception, judgement, and subsequent action.

\begin{figure}[htbp]
    \centering
    \includegraphics[width=0.95\linewidth]{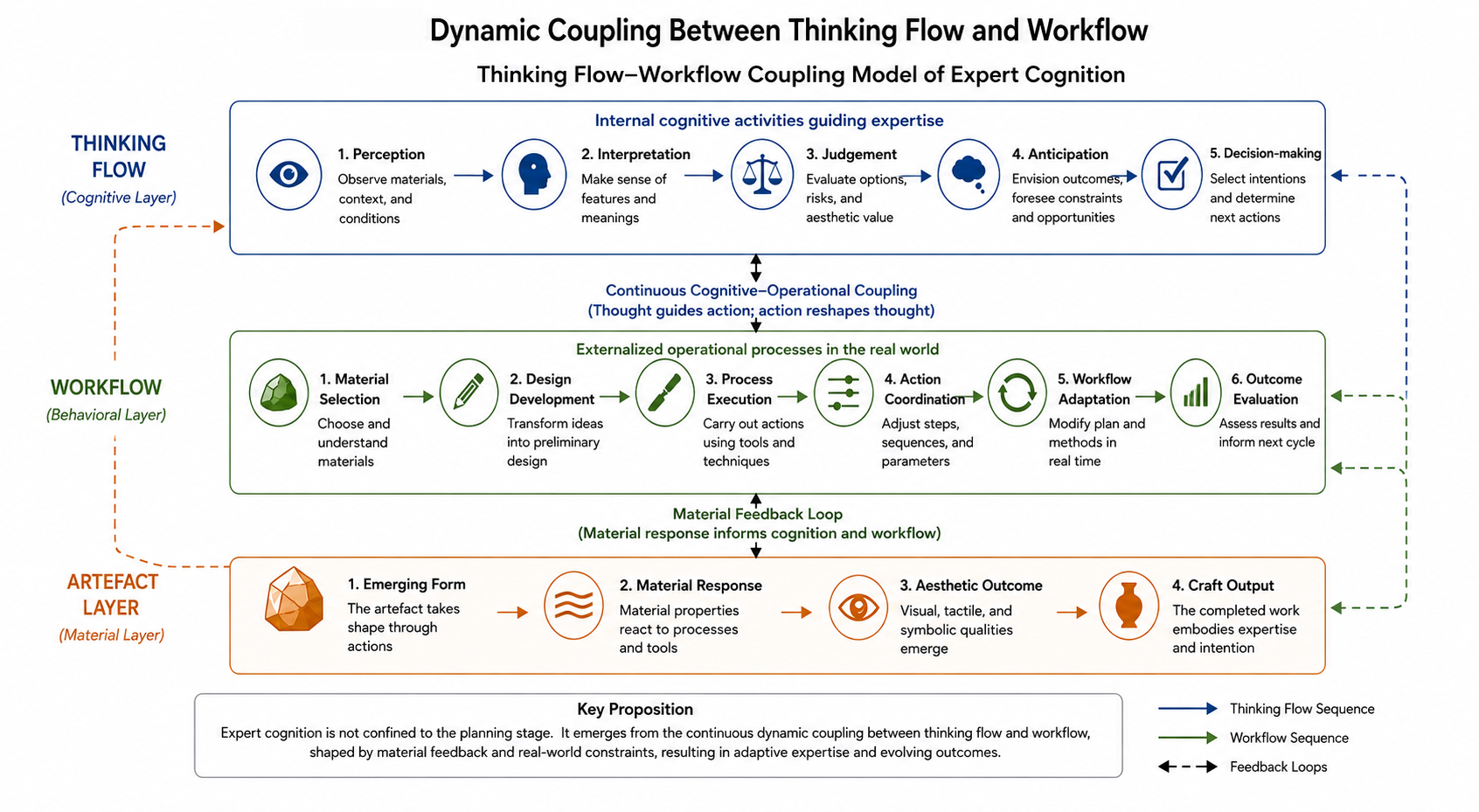}
    \caption{Dynamic coupling between thinking flow, workflow, and artefact in Workflow Cognition.}
    \label{fig:dynamic-coupling}
\end{figure}

As shown in Figure~\ref{fig:dynamic-coupling}, expert practice does not move in a simple linear sequence from thought to action to outcome. Thinking flow guides workflow; workflow transforms the artefact; and the evolving artefact provides material and aesthetic feedback that reshapes thinking flow. Expertise therefore emerges through recursive interaction between cognition, behaviour, and material transformation.

For example, in jade carving, an expert may begin with an initial design intention after examining the colour, translucency, cracks, and internal structure of the stone. This cognitive interpretation guides carving actions. As carving proceeds, newly revealed material features may alter the expert's interpretation of what the object can become. The workflow changes because the thinking flow changes; the thinking flow changes because the material artefact reveals new information. Expertise lies in this continuous adjustment between perception, judgement, action, and material feedback.

\subsection{Workflow Cognition as Expert Practice}

Workflow Cognition provides a way to describe expertise as a dynamic system rather than a collection of procedures. It brings together several dimensions of expert practice.

First, Workflow Cognition is perceptual. Experts attend to features that novices may not notice, such as material variation, compositional imbalance, subtle errors, or emerging opportunities. Second, it is interpretive. Experts make sense of these features in relation to cultural traditions, aesthetic conventions, task goals, and contextual constraints. Third, it is evaluative. Experts judge whether a form, gesture, decision, or outcome is appropriate within the practice. Fourth, it is adaptive. Experts modify actions and workflows in response to changing conditions.

This view aligns with broader accounts of expertise as pattern recognition, situated judgement, reflective action, and adaptive decision-making \cite{schon2017reflective, klein2017sources, ericsson2018cambridge}. It also extends cognitive apprenticeship by specifying what must be made visible if learners are to participate in expert cognition: not only expert actions, but the perceptual cues, interpretive frames, evaluative criteria, adaptive decisions, and material feedback loops that organise those actions.

In ICH education, Workflow Cognition is especially important because aesthetic judgement is embedded within the flow of practice. The expert's judgement is not applied only after the work is complete; it continuously shapes the workflow as the practice unfolds. This means that teaching aesthetic expertise requires more than explaining finished artefacts or demonstrating procedures. It requires making the evolving relationship between perception, judgement, action, and material transformation available to learners.

\subsection{Workflow Evolution and Expertise Formation}

One of the central implications of Workflow Cognition is that expertise should be understood as workflow evolution rather than knowledge accumulation alone. As practitioners repeatedly engage with materials, tasks, environments, and cultural communities, their thinking flows, workflows, and artefact interpretations undergo continuous refinement. They learn what to notice, how to interpret what they notice, when to intervene, and how to adapt when conditions change.

This process connects Workflow Cognition to theories of situated learning and deliberate practice. Learners develop expertise not by acquiring abstract knowledge alone, but by participating in meaningful activity, receiving feedback, and progressively refining their judgement within a domain \cite{lave1991situated, ericsson1993role}. Over time, repeated engagement stabilises certain perceptual and evaluative patterns while also expanding the learner's capacity to respond flexibly to new situations.

In heritage practice, workflow evolution is both technical and aesthetic. It includes improvements in skill, efficiency, and procedural control, but also deeper changes in aesthetic sensitivity, cultural interpretation, and material understanding. A novice may initially ask what step comes next; an expert asks what the material, form, and cultural context now call for. This difference marks the transition from procedural participation to aesthetic cognition.

Workflow Cognition therefore provides the ontological foundation for the framework proposed in this paper. It explains how aesthetic cognition operates within practice and how expert judgement becomes embedded in evolving relationships between cognition, behaviour, and artefact transformation. The next section builds on this foundation by asking how Workflow Cognition can be translated into computable educational structures for AI-native learning environments.

\section{Workflow Cognition Translation}

Section 3 introduced Workflow Cognition as an ontology of expertise and argued that expert performance emerges through the dynamic coupling of thinking flow, workflow, and artefact evolution. However, recognising Workflow Cognition as a theoretical construct does not by itself enable scalable education or AI-mediated learning. If aesthetic cognition is to become accessible beyond traditional apprenticeship settings, Workflow Cognition must become representable, decomposable, and computationally usable.

This section introduces \textit{Workflow Cognition Translation} as the central methodological contribution of the paper. Workflow Cognition Translation provides a framework for transforming expert workflow cognition into computable educational representations capable of supporting AI-native learning environments. It serves as the bridge between Workflow Cognition as an ontology of expertise and Aesthetic Cognition Transmission as an educational paradigm.

\subsection{Why Expert Cognition Must Become Computable}

Traditional educational resources primarily represent explicit knowledge. Textbooks, demonstrations, videos, lectures, and procedural guides are effective for transmitting concepts and techniques, yet they often fail to capture the dynamic cognitive structures through which expertise operates in practice. In heritage domains, this limitation is particularly significant because expert performance frequently depends upon tacit perception, aesthetic judgement, contextual interpretation, material sensitivity, and adaptive decision-making.

If AI systems are to participate meaningfully in heritage education, expert cognition must become computationally representable. Computability in this context does not imply reducing expertise to rigid algorithms or assuming that tacit judgement can be fully formalised. Rather, it refers to constructing representations that preserve pedagogically meaningful relationships between perception, interpretation, judgement, action, feedback, and workflow evolution.

Accordingly, this paper defines Workflow Cognition Translation as follows:

\begin{quote}
\textbf{Methodological Definition.} \textit{Workflow Cognition Translation} is the systematic transformation of expert workflow cognition into computable, AI-readable educational structures that preserve relationships between perception, judgement, decision-making, embodied action, and workflow evolution.
\end{quote}

The objective is not to reproduce expertise in its entirety. Rather, Workflow Cognition Translation seeks to make selected dimensions of expert cognition available for educational generation, learner participation, and human--AI collaboration.

\subsection{Workflow Cognition Decomposition}

Workflow Cognition Translation begins with \textit{Workflow Cognition Decomposition}. Decomposition refers to the systematic process of identifying, extracting, and organising the cognitive structures embedded within expert workflows. The goal is not to fragment expertise into isolated procedural steps, but to reveal the cognitive architecture that supports expert performance.

The decomposition process typically involves four stages.

First, expert thinking flows are elicited through observation, reflective protocols, interviews, stimulated recall, and task analysis. These methods reveal how experts perceive situations, interpret constraints, anticipate consequences, and formulate decisions.

Second, workflow structures are mapped. This stage documents the observable sequence of actions, behavioural transitions, procedural rhythms, and material interactions through which tasks are executed.

Third, cognitive events embedded within workflows are identified. Such events may include moments of aesthetic judgement, opportunity recognition, material interpretation, contextual reasoning, error detection, or adaptive modification.

Finally, extracted cognitive structures are organised into reusable representations capable of supporting computational modelling and educational generation.

\begin{figure}[htbp]
    \centering
    \includegraphics[width=0.95\linewidth]{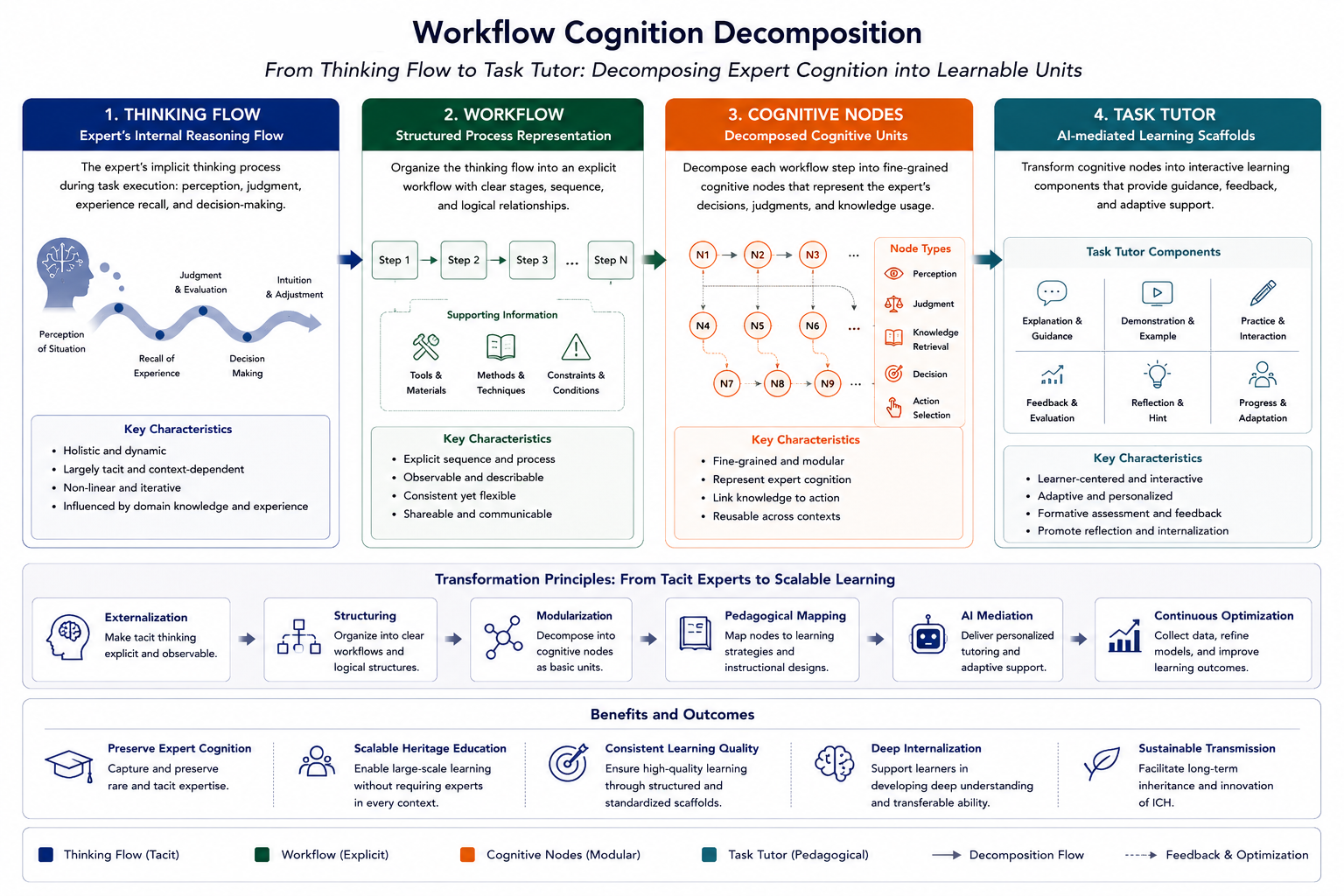}
    \caption{Workflow Cognition Decomposition: From thinking flow and workflow to cognitive nodes and educational generation.}
    \label{fig:workflow-decomposition}
\end{figure}

Figure~\ref{fig:workflow-decomposition} illustrates this process. Expert thinking flows and observable workflows are progressively decomposed into cognitive structures that can subsequently support educational generation.

Through decomposition, Workflow Cognition becomes visible, analysable, and educationally reusable without reducing expertise to static procedural description.

\subsection*{Minimal Cognitive Nodes}

To support computational representation, Workflow Cognition must ultimately be decomposed into smaller operational units. This paper proposes the concept of \textit{Minimal Cognitive Nodes} (MCNs).

\begin{quote}
\textbf{Operational Definition.} A \textit{Minimal Cognitive Node} is the smallest meaningful unit of Workflow Cognition capable of supporting perception, interpretation, judgement, decision-making, behavioural adaptation, or aesthetic evaluation within a workflow context.
\end{quote}

Unlike procedural task steps, Minimal Cognitive Nodes capture the cognitive structures embedded within action.

For example, in jade carving, the instruction remove material from the upper edge'' represents a procedural action. By contrast, the judgement that the colour distribution suggests preserving this region for compositional balance'' represents a cognitive node. The former describes what is done; the latter captures why the action becomes meaningful within the evolving workflow.

Minimal Cognitive Nodes may include perceptual cues, interpretive frames, aesthetic criteria, contextual constraints, decision heuristics, anticipated consequences, or adaptive strategies. Multiple nodes may interact to form larger cognition graphs capable of representing complex expert workflows.

Within educational systems, MCNs provide the building blocks for workflow simulation, adaptive task generation, reflective questioning, and AI-mediated tutoring.

\subsection{Computable Cognition Translation}

Once cognitive structures have been decomposed into Minimal Cognitive Nodes, they may be translated into computable representations.

\textit{Computable Cognition Translation} refers to the process through which Workflow Cognition is encoded into machine-readable structures while preserving relationships between perception, interpretation, judgement, action, and workflow evolution.

The objective is not to replicate human cognition in its entirety. Rather, it is to construct representations capable of supporting educational participation and workflow reasoning.

Possible representations include:

\begin{itemize}
\item cognition graphs;
\item workflow networks;
\item decision pathways;
\item semantic annotations;
\item task structures;
\item behavioural dependencies;
\item aesthetic evaluation frameworks.
\end{itemize}

Through these representations, previously tacit dimensions of expertise become accessible to AI systems without requiring complete formalisation of human cognition. Computable Cognition Translation therefore functions as the bridge between expert practice and AI-native educational environments.

\subsection{AI-Readable Workflow Structures}

Following computational translation, Workflow Cognition may be organised into AI-readable workflow structures.

An \textit{AI-readable workflow structure} is a representation that preserves relationships between thinking flow, workflow actions, cognitive nodes, contextual constraints, and anticipated outcomes. Rather than storing isolated pieces of information, these structures capture the dynamic logic through which expertise operates.

Within AI-mediated educational systems, workflow structures can support:

\begin{itemize}
\item adaptive task generation;
\item workflow simulation;
\item contextual feedback;
\item reflective questioning;
\item personalised learning pathways;
\item learner modelling.
\end{itemize}

Importantly, the educational value of these structures does not depend upon reproducing expert behaviour exactly. Instead, they enable learners to participate in expert reasoning processes and progressively develop aesthetic and procedural understanding.

Educational interaction therefore shifts from content consumption towards cognition participation.

\subsection{Workflow Cognition as Educational Infrastructure}

The ultimate objective of Workflow Cognition Translation is not simply representation but infrastructure.

Once Workflow Cognition has been translated into AI-readable structures, it becomes possible to construct educational systems capable of generating tasks, simulations, learning pathways, and adaptive feedback loops at scale. Workflow Cognition therefore functions as an educational infrastructure layer positioned beneath courses, curricula, and learning activities.

From this perspective, educational content becomes a manifestation of underlying cognition structures rather than an isolated collection of instructional materials. Instead of beginning with courses and then producing content, educational systems may begin with Workflow Cognition itself. Tasks, simulations, reflective activities, and learning pathways can subsequently be generated from translated cognition structures.

This introduces a fundamental shift in educational design. Expertise becomes a reusable educational resource rather than a localised and largely tacit phenomenon. Workflow Cognition Translation therefore provides a pathway through which aesthetic cognition can be transformed into scalable AI-native educational infrastructures capable of supporting broader participation in heritage learning.

The following section examines one application of this framework: AI-native course generation, where translated workflow cognition serves as the foundation for generating adaptive educational experiences.

\section{AI-native Course Generation}

The previous section introduced Workflow Cognition Translation as a methodology for transforming expert workflow cognition into computable educational structures. However, translated cognition structures only become educationally meaningful when they are used to support learning participation. The question therefore becomes how Workflow Cognition can be transformed into educational experiences capable of supporting aesthetic cognition development at scale.

This section proposes \textit{AI-native Course Generation} as a direct application of Workflow Cognition Translation. Rather than treating courses as static collections of instructional content, the framework conceptualises courses as dynamic manifestations of workflow cognition. Educational experiences are generated from translated cognition structures, allowing learners to participate in expert reasoning processes rather than merely consume information about heritage practices.

The approach builds upon traditions of cognitive apprenticeship, learning design, intelligent tutoring systems, and distributed cognition \cite{collins2018cognitive,woolf2010building,laurillard2013teaching,pea1993practices,salomon1997distributed}. However, unlike conventional approaches that begin with curriculum content or instructional objectives, AI-native Course Generation begins with expert workflow cognition itself.

\subsection{From Expert Cases to Educational Structures}

Traditional educational design often begins with curriculum planning, learning outcomes, and content organisation. Educational materials are subsequently created and delivered to learners. Within this model, expertise is typically translated into explanations, demonstrations, and assessments before becoming educational content.

The present framework reverses this process. Instead of beginning with courses, it begins with expert practice. Real-world expert cases become the primary source of educational generation. Through Workflow Cognition Translation, expert activities are transformed into cognition structures capable of supporting learning participation.

Educational structures therefore emerge from expertise rather than being designed independently of it. Expert cases function as educational blueprints from which learning activities, simulations, reflective exercises, and adaptive pathways can be generated. The objective is not to reproduce individual actions, but to reveal the cognitive pathways through which expertise is enacted and refined.

This perspective aligns with recent discussions of AI-native educational systems that move beyond content delivery towards the generation of adaptive learning experiences \cite{luckin2018machine,holmes2019artificial}. However, the present framework extends these ideas by positioning workflow cognition rather than curriculum content as the primary unit of educational generation.

\subsection*{Expert Workflow Extraction}

The first stage of AI-native Course Generation involves expert workflow extraction.

Workflow extraction seeks to identify the operational, cognitive, aesthetic, and contextual structures embedded within expert practice. Data sources may include expert interviews, workflow observation, reflective protocols, process documentation, multimodal recordings, artefact analysis, and community narratives.

The objective is to reconstruct the relationship between thinking flow, workflow, and artefact evolution across the lifecycle of a task.

For example, within jade carving, workflow extraction may reveal a sequence of cognitive and behavioural processes including:

\begin{itemize}
\item material observation;
\item colour interpretation;
\item compositional planning;
\item tool selection;
\item carving decisions;
\item workflow adaptation;
\item aesthetic evaluation.
\end{itemize}

Each stage contains both observable actions and embedded cognitive structures. Workflow extraction therefore provides the empirical foundation for Workflow Cognition Translation.

Importantly, the resulting representation captures more than procedural knowledge. It seeks to represent the perceptual, interpretive, and evaluative dimensions of expertise that guide action throughout practice.

\subsection*{AI-native Task Generation}

Once workflow cognition has been translated into AI-readable structures, educational tasks can be generated directly from the underlying cognition representations.

Unlike conventional educational exercises that primarily assess procedural knowledge or factual recall, AI-native tasks seek to engage learners in expert reasoning processes.

Examples include:

\begin{itemize}
\item aesthetic judgement tasks;
\item workflow decision simulations;
\item material interpretation exercises;
\item comparative reasoning activities;
\item reflective questioning;
\item alternative solution generation;
\item expert-novice comparison tasks.
\end{itemize}

The purpose of these activities is not simply to test whether learners know the correct answer. Rather, they are designed to support participation in expert cognition by exposing learners to the perceptual cues, evaluative criteria, and decision-making structures that shape practice.

For example, a learner may be presented with multiple possible carving strategies and asked to justify a decision based on material characteristics and aesthetic objectives. In such cases, the educational value lies not in selecting the correct procedure, but in engaging with the reasoning processes through which expert decisions are made.

Task generation therefore shifts educational interaction from procedural reproduction towards cognition participation.

\subsection*{Generalised Courseware Generation}

A key advantage of Workflow Cognition Translation is that translated cognition structures can support the generation of multiple forms of educational content.

Because Workflow Cognition is represented through cognitive nodes, workflow graphs, and AI-readable structures, educational systems can generate different learning experiences from the same underlying cognition representation.

A single workflow cognition model may support:

\begin{itemize}
\item introductory heritage learning;
\item workshop preparation programmes;
\item advanced expert training;
\item reflective learning modules;
\item immersive simulations;
\item self-directed learning pathways.
\end{itemize}

This capability enables educational generation at a scale that would be difficult to achieve through manual content production alone. Rather than requiring experts to author complete courses for every audience and context, translated cognition structures become reusable educational assets capable of supporting diverse forms of learning participation.

In this sense, educational production shifts from content authoring towards cognition regeneration.

\subsection{Cross-domain Workflow Transfer}

Although the examples used throughout this paper are drawn primarily from Intangible Cultural Heritage, Workflow Cognition is not inherently domain-specific. Many forms of expertise share common cognitive structures despite substantial differences in tools, materials, environments, and cultural contexts. These shared structures include perception, anticipation, adaptive decision-making, embodied adjustment, aesthetic evaluation, and response to feedback.

For example, in jade carving, Workflow Cognition may involve material perception, colour interpretation, compositional anticipation, adaptive workflow modification, and aesthetic judgement under uncertainty. The expert does not simply apply a fixed carving procedure, but continuously interprets the stone and adapts the workflow according to emerging material conditions.

In wood carving, the tools and materials differ, yet similar cognitive structures may appear. The practitioner must interpret grain direction, structural resistance, form emergence, tool pressure, and surface qualities. Expertise again depends upon the coupling of perception, action, and material feedback. Although jade and wood carving are distinct practices, both require experts to transform material interpretation into adaptive workflow decisions.

Tai Chi provides a different example because it is not organised around artefact production in the same way as craft practices. Nevertheless, expertise formation similarly involves embodied perception, continuous adjustment, procedural rhythm, anticipatory awareness, and longitudinal refinement. The ``artefact'' in this case may be understood less as a material object and more as an evolving embodied form: posture, balance, movement quality, timing, and relational responsiveness. This suggests that Workflow Cognition may extend beyond craft production to broader domains of embodied and aesthetic expertise.

These examples indicate that Workflow Cognition may provide a transferable educational foundation across multiple domains of expertise. The goal is not to claim that all domains share identical cognitive structures, but to suggest that different practices may be represented through comparable relationships between perception, judgement, action, feedback, and adaptation.

The significance of this observation extends beyond heritage education. If Workflow Cognition can be represented through reusable cognitive structures, AI-native educational systems may eventually support the translation of expertise across a wide range of professional, creative, cultural, and embodied domains. Accordingly, the contribution of Workflow Cognition Translation lies not only in supporting ICH education, but also in providing a generalisable framework for transforming expert cognition into educationally usable representations.

\subsection{The Workflow Cognition Translation Loop}

Workflow Cognition Translation should not be understood as a one-directional process.

As learners interact with AI-generated educational systems, they generate new workflow records, reflective responses, decision traces, and participation data. These interactions create opportunities for translated cognition structures to be refined, extended, and validated over time.

Consequently, Workflow Cognition Translation forms a recursive loop consisting of:

\begin{enumerate}
\item expert practice;
\item workflow extraction;
\item cognition translation;
\item AI-native task generation;
\item learner participation;
\item feedback accumulation;
\item cognition refinement.
\end{enumerate}

Rather than preserving expertise as a static archive, the framework supports the development of evolving educational systems capable of adaptation and regeneration.

\begin{figure}[htbp]
    \centering
    \includegraphics[width=0.95\linewidth]{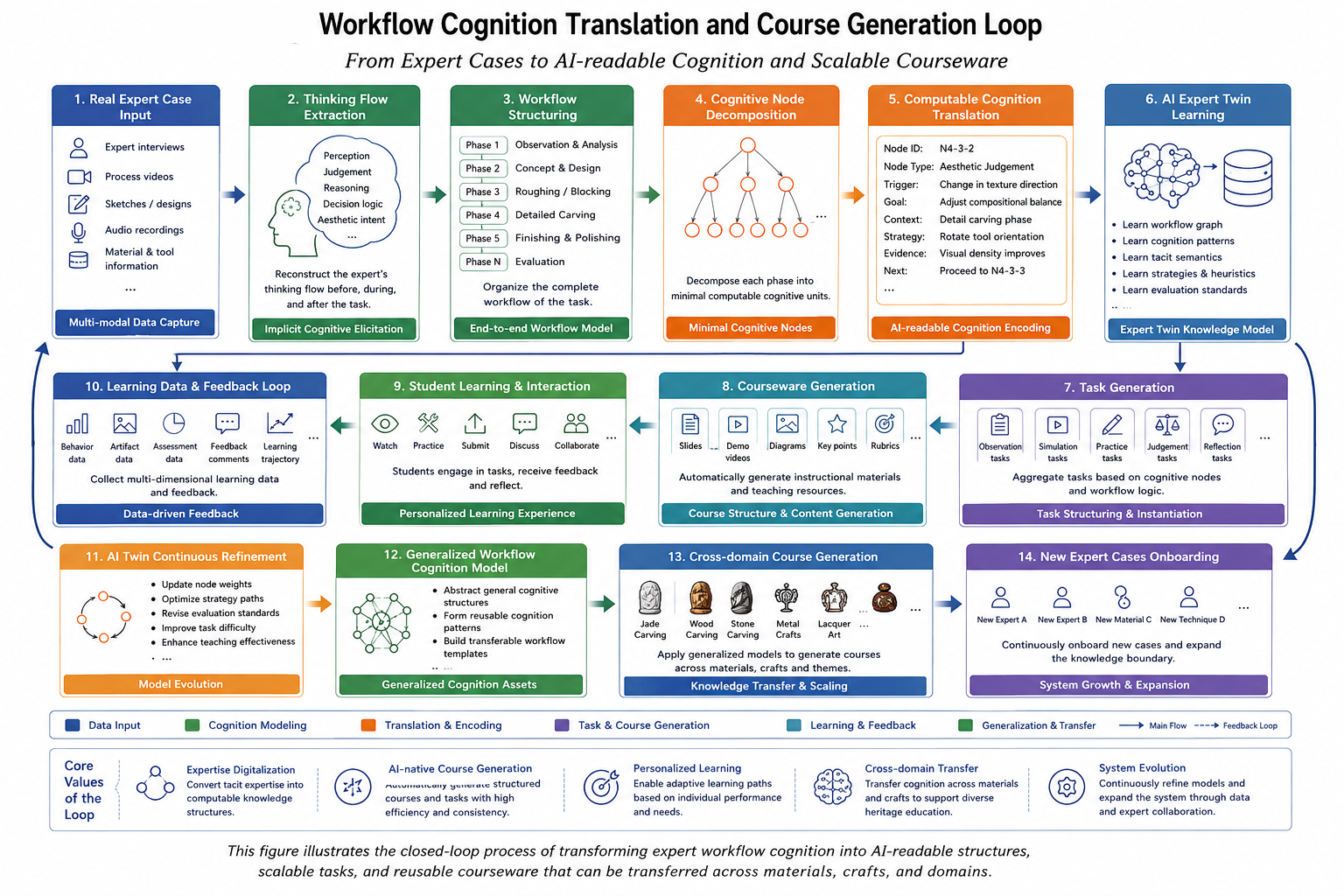}
    \caption{Workflow Cognition Translation and Course Generation Loop: From expert cases to AI-readable cognition and scalable courseware.}
    \label{fig:course-generation}
\end{figure}

Figure~\ref{fig:course-generation} illustrates this process. Workflow Cognition Translation transforms expert cases into AI-readable cognition structures that support task generation, learner participation, and continuous educational refinement.

Viewed in this way, AI-native Course Generation represents not a replacement for traditional educational design, but a new mechanism for constructing learning experiences from expert cognition itself. The following section extends this idea further by proposing an AI-native Cognitive Apprenticeship Infrastructure capable of supporting long-term participation in aesthetic cognition and heritage expertise.

\section{AI-native Cognitive Apprenticeship Infrastructure}

The preceding sections introduced Aesthetic Cognition Transmission as an educational paradigm, Workflow Cognition as an ontology of expertise, and Workflow Cognition Translation as a methodology for transforming expert cognition into computable educational structures. Together, these concepts establish the foundations for AI-native heritage education. However, educational generation alone does not constitute a complete learning ecosystem. Scalable participation in aesthetic cognition requires an infrastructure capable of preserving expertise, supporting learner interaction, and enabling long-term cognition development.

This section proposes an \textit{AI-native Cognitive Apprenticeship Infrastructure}. Rather than organising educational systems around content delivery, the proposed infrastructure is organised around cognition representation, workflow participation, adaptive task generation, and longitudinal expertise development. The objective is not to replace apprenticeship, but to extend access to expert cognitive pathways through AI-mediated participation systems.

\subsection{The Longitudinal Tacit Cognition Infrastructure Model}

\begin{figure}[htbp]
    \centering
    \includegraphics[width=0.95\linewidth]{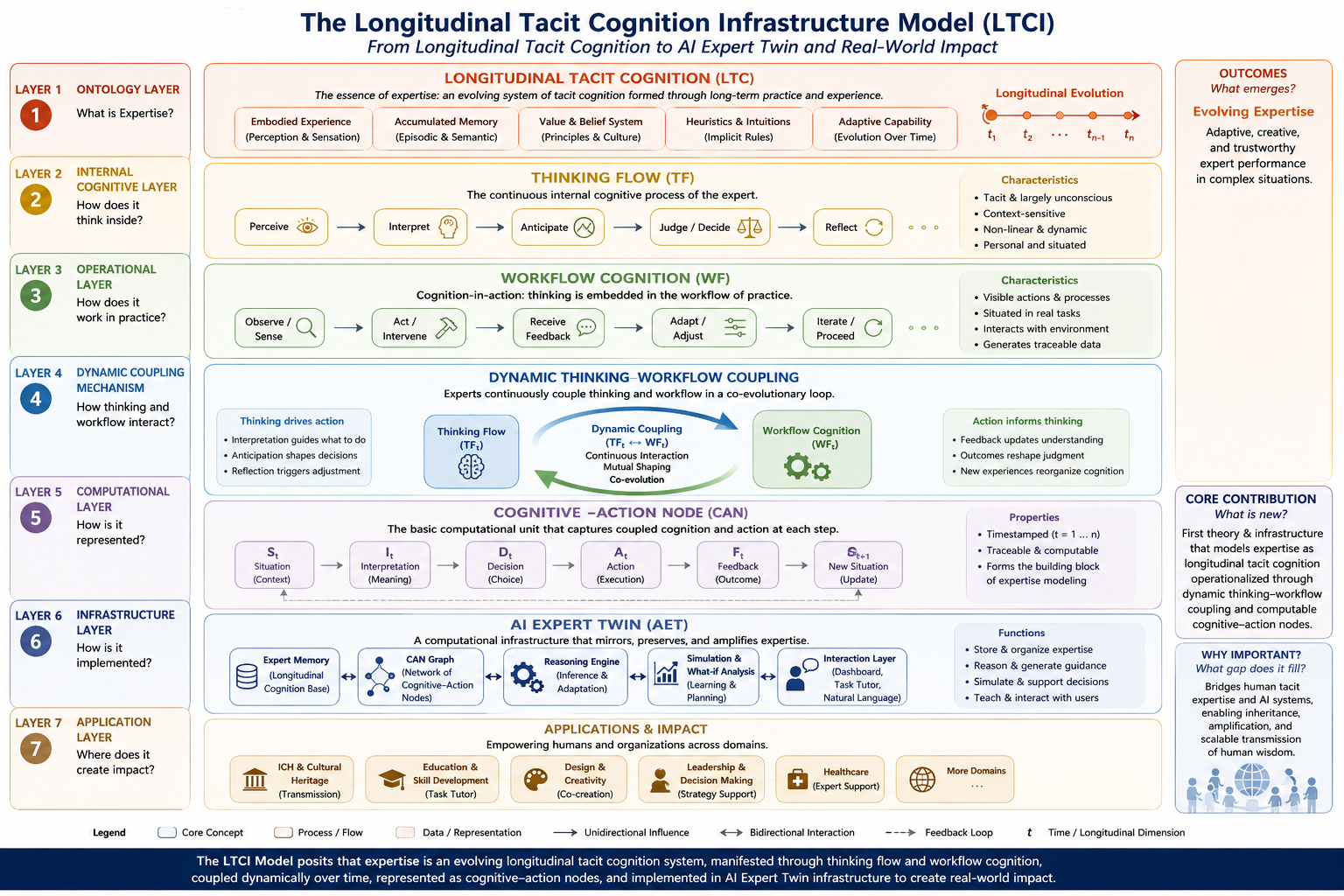}
    \caption{The Longitudinal Tacit Cognition Infrastructure (LTCI) model: from longitudinal tacit cognition to AI Expert Twins and real-world impact.}
    \label{fig:ltci}
\end{figure}

The framework developed throughout this paper can be synthesised through the \textit{Longitudinal Tacit Cognition Infrastructure} (LTCI) model shown in Figure~\ref{fig:ltci}. The LTCI model conceptualises expertise as a multi-layered system that extends from the ontological foundations of expertise to AI-mediated educational infrastructures and real-world impact.

The first layer is the \textit{Ontology Layer}, which addresses the question: \emph{What is expertise?} Within the present framework, expertise is understood as longitudinal tacit cognition that develops through extended participation in practice, cultural environments, and material engagement.

The second layer is the \textit{Internal Cognitive Layer}, which addresses the question: \emph{How does expertise think?} This layer corresponds to thinking flow: the ongoing cognitive processes through which experts perceive situations, interpret context, anticipate possibilities, evaluate alternatives, and generate decisions.

The third layer is the \textit{Operational Layer}, which addresses the question: \emph{How does expertise operate in practice?} This layer corresponds to Workflow Cognition, where thinking becomes embedded within workflow execution, material interaction, and situated action.

The fourth layer is the \textit{Dynamic Coupling Layer}, which addresses the relationship between thinking flow and workflow. Expertise emerges not from cognition or action in isolation, but from their continuous co-evolution. Experts constantly adapt thinking in response to workflow feedback and simultaneously modify workflows based on evolving judgement and interpretation.

The fifth layer is the \textit{Computational Layer}, which addresses the question: \emph{How can expertise be represented computationally?} Within this paper, Workflow Cognition Translation and computational cognition structures provide the mechanisms through which expert cognition becomes representable and educationally usable.

The sixth layer is the \textit{Infrastructure Layer}, which addresses the question: \emph{How can expertise be implemented and sustained?} This layer introduces AI-native systems such as AI Expert Twins, Task Tutors, and learner support agents capable of mediating access to workflow cognition.

Finally, the seventh layer is the \textit{Application Layer}, which addresses the question: \emph{Where does expertise create impact?} Through educational, organisational, cultural, and professional applications, translated cognition structures may support broader participation in expertise and contribute to long-term human development.

The LTCI model therefore provides a unifying framework that connects the theoretical, methodological, computational, and educational components introduced throughout this paper.

\subsection*{AI Expert Twins}

At the centre of the proposed infrastructure is the concept of the \textit{AI Expert Twin}.

An AI Expert Twin is not simply a digital archive, knowledge base, or conversational agent. Rather, it is a computational representation of Workflow Cognition constructed through Workflow Cognition Translation. Its purpose is to preserve and mediate access to expert cognitive structures rather than merely storing information about a domain.

An AI Expert Twin may incorporate:

\begin{itemize}
\item workflow structures;
\item cognitive nodes;
\item decision pathways;
\item aesthetic judgement patterns;
\item tacit semantic systems;
\item behavioural rhythms;
\item longitudinal domain experience.
\end{itemize}

The primary function of the AI Expert Twin is not to reproduce expert behaviour exactly, but to provide educational access to expert cognition. Through interaction with learners, the system may explain workflow decisions, generate adaptive learning tasks, simulate expert reasoning, provide contextual feedback, and support reflective learning.

In this sense, the AI Expert Twin functions as a cognition mediation system rather than a knowledge retrieval system.

\subsection*{Task Tutors and Workflow Participation}

Building upon the AI Expert Twin, the proposed infrastructure introduces \textit{Task Tutors}. Whereas the AI Expert Twin represents a broader expertise model, Task Tutors focus on specific workflows, projects, or learning objectives.

Task Tutors may:

\begin{itemize}
\item guide workflow participation;
\item scaffold task performance;
\item simulate expert decision-making;
\item evaluate learner responses;
\item generate reflective prompts;
\item support workflow exploration.
\end{itemize}

Importantly, Task Tutors are not conceived as instructional assistants that merely deliver content. Rather, they function as workflow cognition interfaces through which learners engage directly with translated expertise.

This reflects a broader shift from content-centred education towards cognition-centred participation. Learners are not only exposed to information about a domain, but are progressively guided into expert modes of perception, interpretation, and decision-making.

\subsection*{Student Tutors and Longitudinal Learning}

Whilst Task Tutors support specific learning activities, \textit{Student Tutors} support long-term learner development.

Student Tutors function as persistent learner-facing agents that accumulate information concerning learner participation, workflow experience, reflective records, educational trajectories, and developing forms of aesthetic understanding.

Unlike conventional learning management systems that primarily track completion and performance metrics, Student Tutors are designed to model cognitive development over time.

Such systems may support:

\begin{itemize}
\item personalised learning pathways;
\item adaptive workflow recommendations;
\item cognition gap identification;
\item reflective learning support;
\item longitudinal aesthetic development.
\end{itemize}

This capability aligns closely with the paper's emphasis on aesthetic cognition as a developmental process rather than a collection of isolated competencies. Learning becomes a process of long-term participation in cognition systems rather than progression through disconnected instructional modules.

\subsection*{Human--AI Cognitive Collaboration}

The proposed infrastructure does not conceptualise AI as an autonomous educational authority. Instead, expertise development emerges through collaboration between experts, learners, AI systems, workflows, and cultural environments.

Within this model:

\begin{itemize}
\item experts contribute workflow cognition and cultural knowledge;
\item AI systems translate, organise, and mediate cognitive structures;
\item learners participate in workflow experiences and generate new learning data;
\item communities provide cultural context, validation, and continuity.
\end{itemize}

This perspective aligns with broader theories of distributed cognition \cite{pea1993practices,salomon1997distributed,hutchins1995cognition}. Expertise is not located solely within individuals but emerges through interactions between people, artefacts, environments, and computational systems.

AI therefore functions as a cognition infrastructure that expands access to expertise rather than replacing human expertise itself.

\subsection{Towards AI-native Aesthetic Cognition Ecosystems}

The concepts introduced throughout this paper ultimately converge towards a broader vision of AI-native aesthetic cognition ecosystems.

Traditional heritage education has historically been organised around workshops, local communities, and apprenticeship relationships. These structures remain essential for preserving cultural authenticity and embodied practice. However, they are difficult to scale and often inaccessible to wider audiences.

The infrastructure proposed here introduces an additional layer of participation.

Within this ecosystem:

\begin{itemize}
\item cultural aesthetics provide the source of expertise;
\item Workflow Cognition provides an ontology of expertise;
\item Workflow Cognition Translation provides a computational methodology;
\item AI Expert Twins provide cognition representation;
\item Task Tutors provide workflow participation;
\item Student Tutors provide longitudinal learning support;
\item learners contribute interaction, reflection, and feedback;
\item AI systems continuously refine educational structures.
\end{itemize}

The result is an educational ecosystem organised around participation in cognition rather than the delivery of content. Heritage education becomes a living infrastructure for preserving, translating, and transmitting aesthetic cognition across time.

This perspective extends the role of AI beyond documentation, retrieval, or automation. Instead, AI becomes part of a broader infrastructure for sustaining human expertise, cultural participation, and aesthetic understanding. In this sense, AI-native Cognitive Apprenticeship Infrastructure represents not merely a technological system, but a framework for expanding access to expertise while preserving the cultural, experiential, and human foundations of heritage practice.

\section{AI-mediated ICH Learning Pathway}

The previous section proposed an AI-native Cognitive Apprenticeship Infrastructure for preserving, translating, and mediating access to expert cognition. However, infrastructure alone does not explain how learners progressively enter heritage practice. This section therefore proposes an \textit{AI-mediated ICH Learning Pathway}: a five-stage model through which learners move from initial aesthetic exposure towards deeper participation in expert workflow cognition and, eventually, situated workshop practice.

The pathway is designed to clarify the educational role of AI-mediated systems in ICH learning. It does not replace apprenticeship, material practice, or cultural immersion. Rather, it provides preparatory and complementary forms of participation that can make expert cognition more accessible before learners enter full workshop environments.

\begin{figure}[htbp]
    \centering
    \includegraphics[width=0.95\linewidth]{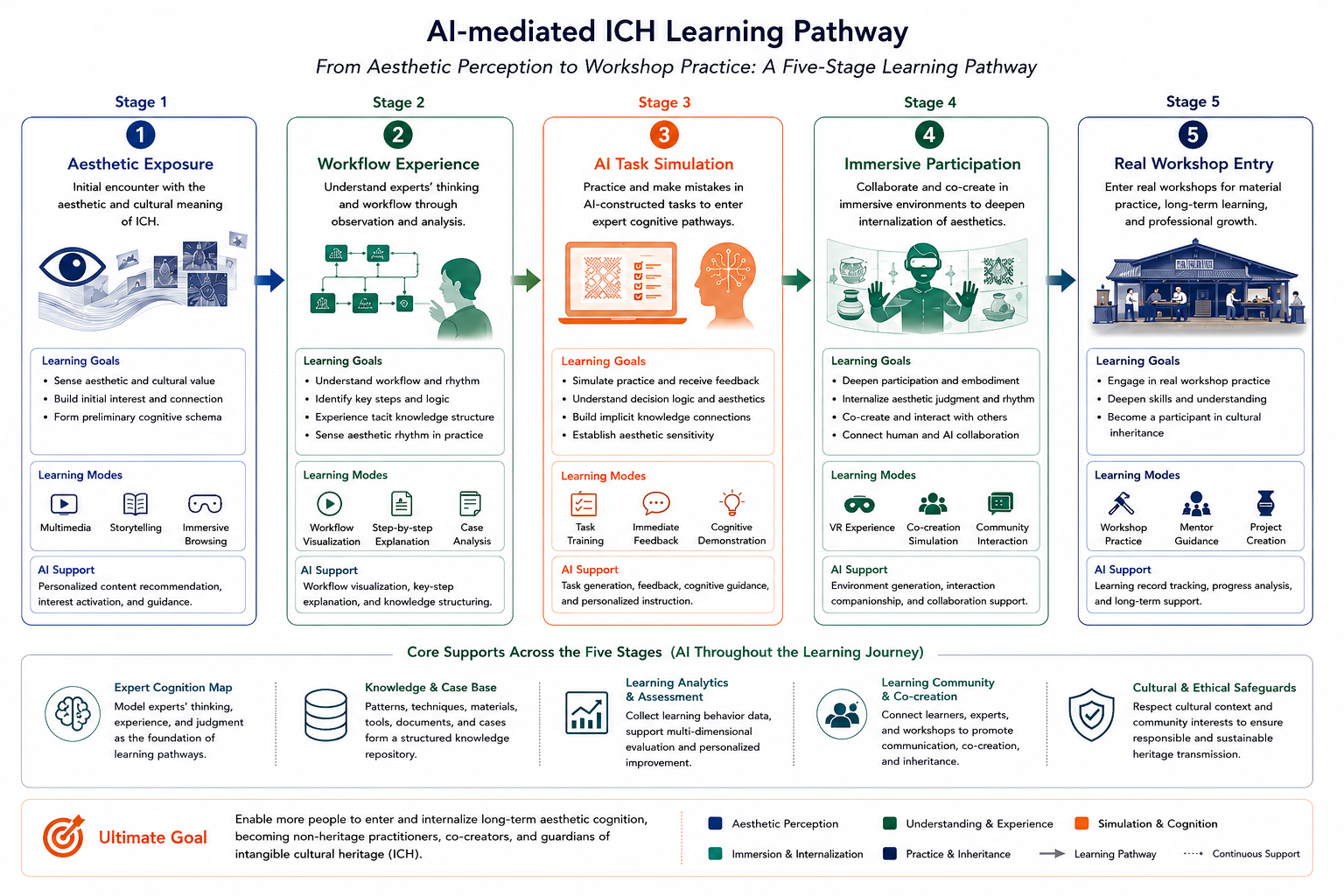}
    \caption{AI-mediated ICH Learning Pathway: From aesthetic perception to workshop practice.}
    \label{fig:ich-learning-pathway}
\end{figure}

As shown in Figure~\ref{fig:ich-learning-pathway}, the proposed pathway consists of five stages: aesthetic exposure, workflow experience, AI task imitation, immersive participation, and real workshop entry.

\subsection*{Stage 1: Aesthetic Exposure}

The first stage is \textit{aesthetic exposure}. At this stage, learners encounter the visual forms, symbolic meanings, cultural narratives, historical significance, and aesthetic qualities of a heritage practice. The goal is not yet technical competence, but aesthetic attraction, cultural curiosity, and initial orientation.

Historically, such exposure often occurred through family traditions, local communities, workshops, performances, or direct contact with practitioners. In contemporary settings, however, learners may first encounter ICH through digital exhibitions, online media, virtual collections, cultural platforms, or AI-mediated learning environments.

Aesthetic exposure therefore functions as the entry point into heritage learning. It allows learners to begin noticing aesthetic forms and cultural meanings before they are expected to reproduce techniques or participate in complex workflows.

\subsection*{Stage 2: Workflow Experience}

The second stage is \textit{workflow experience}. Here, learners begin to understand how heritage practice unfolds through expert perception, judgement, decision-making, and action.

Rather than focusing only on finished artefacts, learners are introduced to the process through which expert work develops. They may observe workflow structures, process narratives, expert decision pathways, material transformations, and behavioural rhythms.

For example, in jade carving, learners may examine how an expert interprets colour distribution, evaluates structural constraints, develops a compositional plan, selects tools, and modifies decisions as the material changes. The purpose is to reveal heritage practice as an evolving cognitive process rather than a fixed sequence of procedures.

Workflow experience therefore prepares learners to see practice as cognition in action.

\subsection*{Stage 3: AI Task Imitation}

The third stage is \textit{AI task imitation}. At this stage, learners engage with AI-constructed tasks generated from translated workflow cognition structures. These tasks allow learners to practise judgement, make mistakes, compare alternatives, and rehearse expert-like reasoning in a low-risk environment.

AI task imitation may include aesthetic judgement exercises, workflow decision simulations, material interpretation tasks, comparative reasoning activities, and reflective prompts. Learners may be asked to choose between alternative actions, justify design decisions, identify errors, or explain why a particular workflow adaptation is appropriate.

The purpose of imitation here is not mechanical copying. Rather, it is structured entry into expert cognitive pathways. By engaging with AI-generated tasks, learners begin to internalise the perceptual cues, evaluative criteria, and decision structures that organise expert practice.

\subsection*{Stage 4: Immersive Participation}

The fourth stage is \textit{immersive participation}. At this stage, learners move beyond discrete AI-generated tasks towards richer forms of interaction, collaboration, and co-creation within immersive environments.

Immersive participation may involve virtual reality, augmented reality, mixed reality, interactive digital workshops, collaborative simulations, or AI-supported design environments. Such environments allow learners to experience workflows dynamically, explore consequences, participate in shared activities, and deepen their internalisation of aesthetic structures.

The educational aim is to strengthen the relationship between cognition, embodiment, and cultural meaning. Learners do not merely observe expert practice or complete isolated tasks; they begin to participate in environments that approximate the complexity, rhythm, and social dimensions of heritage learning.

\subsection*{Stage 5: Real Workshop Entry}

The final stage is \textit{real workshop entry}. Despite the potential of AI-mediated systems, heritage expertise ultimately remains connected to material practice, embodied action, social apprenticeship, and situated cultural participation.

Real workshops provide forms of experience that cannot be fully replicated digitally, including tactile material interaction, tool handling, embodied motor development, environmental awareness, social learning, and long-term mentorship. For this reason, the proposed pathway does not treat AI-mediated learning as a substitute for workshop practice.

Instead, AI-mediated learning prepares learners for deeper participation. By the time learners enter physical workshop environments, they may already have developed initial aesthetic awareness, workflow understanding, expert reasoning patterns, and reflective vocabulary. Workshop entry therefore becomes not the beginning of learning, but the continuation of a longer cognitive apprenticeship pathway.

\subsection{Implications for Scalable Heritage Participation}

The five-stage pathway expands the meaning of participation in ICH education. Learners do not need to begin immediately with full craft production or professional apprenticeship. Instead, they may progressively enter heritage cognition through exposure, observation, simulation, immersion, and eventually embodied practice.

This model also clarifies the role of AI within heritage education. AI is most valuable not when it replaces masters or workshops, but when it supports earlier and broader forms of access to aesthetic cognition. It can help learners notice what experts notice, reason through workflow decisions, practise judgement, and prepare for more meaningful material engagement.

In this sense, the AI-mediated ICH Learning Pathway provides a bridge between scalable digital participation and situated heritage practice. It allows heritage education to expand access while preserving the central importance of human expertise, workshop environments, and long-term cultural participation.

\section{Conclusion}

This paper has argued that the future of scalable Intangible Cultural Heritage (ICH) education may depend not simply on preserving crafts, documenting artefacts, or digitising procedural knowledge, but on expanding access to the cognitive structures through which heritage expertise is developed and enacted. While traditional apprenticeship remains indispensable for cultivating advanced mastery, contemporary digital technologies create new opportunities for broader participation in heritage learning. The central challenge is therefore not only how to preserve heritage practices, but how to preserve and transmit the aesthetic cognition embedded within them.

To address this challenge, the paper proposed \textit{Aesthetic Cognition Transmission} as an educational paradigm that shifts the focus of heritage education from craft reproduction towards participation in expert ways of perceiving, interpreting, judging, and acting. Building on theories of aesthetic education, tacit knowledge, situated learning, expertise, and cognitive apprenticeship, the paper argued that many of the most valuable dimensions of heritage practice reside not solely in techniques or artefacts, but in the cognitive processes that guide expert action.

The paper introduced \textit{Workflow Cognition} as an ontology of expertise capable of describing how perception, judgement, decision-making, embodied action, and adaptation operate within evolving workflows. It further proposed \textit{Workflow Cognition Translation} as a methodological framework for transforming expert workflow cognition into computable educational structures. Through Workflow Cognition Decomposition, Minimal Cognitive Nodes, and AI-readable workflow structures, previously tacit dimensions of expertise become accessible to AI-native educational systems.

Building upon this foundation, the paper outlined a framework for AI-native Course Generation, AI-native Cognitive Apprenticeship Infrastructure, and AI-mediated ICH Learning Pathways. These components collectively demonstrate how translated workflow cognition may support scalable participation in aesthetic expertise while preserving the importance of workshops, embodied practice, and human mentorship. Within this framework, AI functions not as a replacement for masters or apprenticeship traditions, but as a cognition mediation infrastructure that expands access to expert cognitive pathways.

Beyond heritage education, the concepts developed in this paper may have broader implications for the future of AI-native learning systems. Many forms of expertise, including craftsmanship, design, healthcare, leadership, and professional practice, rely upon tacit judgement, workflow reasoning, and long-term cognitive development. Workflow Cognition Translation therefore suggests a more general approach to educational design in which expert cognition becomes a reusable educational resource rather than remaining confined to localised communities of practice.

Ultimately, the contribution of this paper lies not in proposing a new educational technology, but in offering a new way of thinking about expertise, learning, and cultural transmission in the AI era. If the twentieth century focused on preserving information and the early twenty-first century focused on preserving digital artefacts, the next challenge may be preserving and transmitting human cognition itself. From this perspective, the future of heritage education is not only about safeguarding cultural objects or craft techniques, but about enabling broader participation in the aesthetic, cognitive, and cultural foundations through which human expertise is formed and sustained.

\subsection*{Copyright Notice}

\noindent
\textcopyright\ 2026 Annie Yihong Yuan. All rights reserved. AAll figures, diagrams, interface examples, and visual materials in this paper are original works of the author unless otherwise stated.


\bibliographystyle{unsrtnat}
\bibliography{references}  

@incollection{dewey2024art,
  title={Art as experience},
  author={Dewey, John},
  booktitle={Anthropology of the Arts},
  pages={37--45},
  year={2024},
  publisher={Routledge}
}

@inproceedings{dewey1986experience,
  title={Experience and education},
  author={Dewey, John},
  booktitle={The educational forum},
  volume={50},
  number={3},
  pages={241--252},
  year={1986},
  organization={Taylor \& Francis}
}

@article{eisner2003arts,
  title={The arts and the creation of mind},
  author={Eisner, Elliot W},
  journal={Language arts},
  volume={80},
  number={5},
  pages={340--344},
  year={2003},
  publisher={NCTE}
}

@book{eisner2017enlightened,
  title={The enlightened eye: Qualitative inquiry and the enhancement of educational practice},
  author={Eisner, Elliot W},
  year={2017},
  publisher={Teachers College Press}
}

@book{greene2000releasing,
  title={Releasing the imagination: Essays on education, the arts, and social change},
  author={Greene, Maxine},
  year={2000},
  publisher={John Wiley \& Sons}
}

@book{shusterman2000pragmatist,
  title={Pragmatist aesthetics: Living beauty, rethinking art},
  author={Shusterman, Richard},
  year={2000},
  publisher={Bloomsbury Publishing PLC}
}

@book{shusterman2008body,
  title={Body consciousness: A philosophy of mindfulness and somaesthetics},
  author={Shusterman, Richard},
  year={2008},
  publisher={Cambridge University Press}
}

@incollection{polanyi2009tacit,
  title={The tacit dimension},
  author={Polanyi, Michael},
  booktitle={Knowledge in organisations},
  pages={135--146},
  year={2009},
  publisher={Routledge}
}

@book{lave1991situated,
  title={Situated learning: Legitimate peripheral participation},
  author={Lave, Jean and Wenger, Etienne},
  year={1991},
  publisher={Cambridge university press}
}

@book{wenger1999communities,
  title={Communities of practice: Learning, meaning, and identity},
  author={Wenger, Etienne},
  year={1999},
  publisher={Cambridge university press}
}

@book{schon2017reflective,
  title={The reflective practitioner: How professionals think in action},
  author={Sch{\"o}n, Donald A},
  year={2017},
  publisher={Routledge}
}

@incollection{collins2018cognitive,
  title={Cognitive apprenticeship: Teaching the crafts of reading, writing, and mathematics},
  author={Collins, Allan and Brown, John Seely and Newman, Susan E},
  booktitle={Knowing, learning, and instruction},
  pages={453--494},
  year={2018},
  publisher={Routledge}
}

@article{collins1991cognitive,
  title={Cognitive apprenticeship: Making thinking visible},
  author={Collins, Allan and Brown, John Seely and Holum, Ann and others},
  journal={American educator},
  volume={15},
  number={3},
  pages={6--11},
  year={1991}
}

@article{brown1989situated,
  title={Situated cognition and the culture of learning},
  author={Brown, John Seely and Collins, Allan and Duguid, Paul},
  journal={Educational researcher},
  volume={18},
  number={1},
  pages={32--42},
  year={1989},
  publisher={Sage Publications Sage CA: Thousand Oaks, CA}
}

@book{rogoff1990apprenticeship,
  title={Apprenticeship in thinking: Cognitive development in social context},
  author={Rogoff, Barbara},
  year={1990},
  publisher={Oxford university press}
}

@book{simon2019sciences,
  title={The Sciences of the Artificial, reissue of the third edition with a new introduction by John Laird},
  author={Simon, Herbert A},
  year={2019},
  publisher={MIT press}
}

@article{ericsson1993role,
  title={The role of deliberate practice in the acquisition of expert performance.},
  author={Ericsson, K Anders and Krampe, Ralf T and Tesch-R{\"o}mer, Clemens},
  journal={Psychological review},
  volume={100},
  number={3},
  pages={363},
  year={1993},
  publisher={American Psychological Association}
}

@book{chi2014nature,
  title={The nature of expertise},
  author={Chi, Michelene TH and Glaser, Robert and Farr, Marshall J},
  year={2014},
  publisher={Psychology Press}
}

@book{woolf2010building,
  title={Building intelligent interactive tutors: Student-centered strategies for revolutionizing e-learning},
  author={Woolf, Beverly Park},
  year={2010},
  publisher={Morgan Kaufmann}
}

@book{luckin2018machine,
  title={Machine Learning and Human Intelligence. The future of education for the 21st century},
  author={Luckin, Rosemary},
  year={2018},
  publisher={UCL institute of education press}
}

@book{holmes2019artificial,
  title={Artificial intelligence in education promises and implications for teaching and learning},
  author={Holmes, Wayne and Bialik, Maya and Fadel, Charles},
  year={2019},
  publisher={Center for Curriculum Redesign}
}

@article{pea1993practices,
  title={Practices of distributed intelligence and designs for education},
  author={Pea, Roy D},
  journal={Distributed cognitions: Psychological and educational considerations},
  volume={11},
  pages={47--87},
  year={1993}
}

@book{salomon1997distributed,
  title={Distributed cognitions: Psychological and educational considerations},
  author={Salomon, Gavriel},
  year={1997},
  publisher={Cambridge University Press}
}

@book{hutchins1995cognition,
  title={Cognition in the Wild},
  author={Hutchins, Edwin},
  year={1995},
  publisher={MIT press}
}

@book{laurillard2013teaching,
  title={Teaching as a design science: Building pedagogical patterns for learning and technology},
  author={Laurillard, Diana},
  year={2013},
  publisher={Routledge}
}

@book{klein2017sources,
  title={Sources of power: How people make decisions},
  author={Klein, Gary A},
  year={2017},
  publisher={MIT press}
}

@book{ericsson2018cambridge,
  title={The Cambridge handbook of expertise and expert performance},
  author={Ericsson, K Anders and Hoffman, Robert R and Kozbelt, Aaron},
  year={2018},
  publisher={Cambridge University Press}
}






\end{document}